\documentclass[journal]{IEEEtran}
\pdfoutput=1
\usepackage{mathbbold}
\usepackage{mathrsfs}

\ifCLASSINFOpdf
\else
\fi
\usepackage{amsmath}
\usepackage{amsfonts}
\usepackage{graphicx}
\usepackage{tabulary}
\usepackage{caption}
\usepackage{amsmath}
\usepackage{color}
\usepackage{cases}
\usepackage{amsfonts,amsmath,amssymb}
\usepackage{mathrsfs}
\usepackage{cite}
\usepackage{graphicx}
\usepackage{url}
\usepackage{enumerate}
\usepackage{subfigure}
\usepackage{hyperref}
\usepackage{algorithm}
\usepackage{algorithmic}
\usepackage{amsmath,bm}
\begin{document}

%

% Titles are generally capitalized except for words
\title{Brain Maturation Study during Adolescence Using Graph Laplacian Learning Based Fourier Transform}

\author{Junqi~Wang, Li Xiao, Tony~W.~Wilson, Julia~M.~Stephen, Vince~D.~Calhoun, Yu-Ping~Wang% <-this % stops a space
\thanks{J. Wang, L.Xiao and Y. Wang were with the Department
of Biomedical Engineering, Tulane University, New Orleans,
LA, USA e-mail: wyp@tulane.edu}% <-this % stops a space
\thanks{V.D Calhoun was with the Department of Electrical and Computer Engineering, The University of New Mexico, Albuquerque, NM, USA and also with the Mind Research Network, Albuquerque.}
\thanks{J.M. Stephen was with the Mind Research Network, Albuquerque, NM, USA}
\thanks{T.W Wilson was with the Department of Neurological Sciences, University of Nebraska Medical Center,Omaha, NE, USA}}
% <-this % stops a space
%\thanks{Manuscript finished in March 15, 2018}}

%\markboth{Happy New Year}%
% The only time the second header will appear is for
\maketitle

\begin{abstract}
Objective: Longitudinal neuroimaging studies have demonstrated that adolescence is the crucial developmental epoch of continued brain growth and change. A large number of researchers dedicate to uncovering the mechanisms about brain maturity during adolescence. Motivated by both achievement in graph signal processing and recent evidence that some brain areas act as hubs connecting functionally specialized systems, we proposed an approach to detect these regions from spectral analysis perspective. In particular, as human brain undergoes substantial development throughout adolescence, we addressed the challenge by evaluating the functional network difference among age groups from functional magnetic resonance imaging (fMRI) observations. \par
Methods: We treated these observations as graph signals defined on the parcellated functional brain regions and applied graph Laplacian learning based Fourier Transform (GLFT) to transform the original graph signals into frequency domain. Eigen-analysis was conducted afterwards to study the behavior of the corresponding brain regions, which enables the characterization of brain maturation. \par
Result: We first evaluated our method on the synthetic data and further applied the method to resting and task state fMRI imaging data from Philadelphia Neurodevelopmental Cohort (PNC) dataset, comprised of normally developing adolescents from 8 to 22. The model provided a highest accuracy of 95.69\% in distinguishing different adolescence stages. \par
Conclusion: We detected 13 hubs from resting state fMRI and 16 hubs from task state fMRI that are highly related to brain maturation process. \par
Significance: The proposed GLFT method is powerful in extracting the brain connectivity patterns and identifying hub regions with a high prediction power.

\end{abstract}

\begin{IEEEkeywords}
brain maturation, GLFT, PNC, fMRI, hubs.
\end{IEEEkeywords}

\IEEEpeerreviewmaketitle

\section{Introduction}
% The very first letter is a 2 line initial drop letter
\IEEEPARstart AdVANCES in neuroimaging techniques for the study of human brain have attracted a growing amount of scientific interest in recent years. In particular, functional magnetic resonance imaging (fMRI) provides a chance to study brain structure and function through non-invasive measurements over a time frame of seconds with a spatial resolution of millimeters. Resting state functional MRI (rs-fMRI), which measures spontaneous low frequency fluctuations in blood oxygenation level dependent (BOLD) signal, has attracted attention for its ability to measure correlations in neural activity between distant brain regions \cite{power2011functional}. These correlations can have multiple applications in improving medical diagnostics, prognostics and further therapeutics. Task state fMRI (tfMRI) measures the behavior of the brain activity with a range of motor, sensory, cognitive and emotional process tasks to explore a core set of functional related brain connectivity and human function. tfMRI data can also help characterize and validate the connectivity analyses conducted on structural and resting-state functional data \cite{barch2013function}.

It is a common assumption that the brain operates as a set of distributed sub-networks \cite{eavani2015identifying} that co-activate over time. A significant amount of studies have been conducted to uncover the mechanisms about changes in brain functional activity that occur throughout childhood and adolescence \cite{jolles2010comprehensive,fair2009functional}. Task state fMRI data have been used to detect the functional behavioral difference between children and adult brain \cite{passarotti2003development}. Interesting findings have drawn wide concerns about exploring the core regions with brain development \cite{holland2001normal}. As the functional organization of the brain structure is still not fully understood, there is a growing interest in capturing the structure of the brain network during early development.
\par
    A variety of methods have been proposed to study the neuroimaging data and one widely used approach is the graph based one \cite{bullmore2009complex,rubinov2010complex}. Within this framework, a complex system is formulated as a graph or network consisting of a set of nodes and pairwise relationships between the nodes, i.e., edges. Spectral graph theory was introduced to study the properties of these graphs via the eigen-analysis of the associated graph matrices: the adjacency matrix, the graph Laplacian and their variants. These matrices have been well studied as the representation of graph signals with many applications \cite{chung1997spectral}, which led to the emerging field of graph signal processing \cite{sandryhaila2013discrete}. Among these graph signal processing tools, the graph Fourier transform (GFT) is a powerful one with well-documented performance in signal denoising \cite{shuman2013emerging}, mental disease study \cite{hu2013matched}, and brain network studies \cite{huang2016graph}. As a generalization of the Fourier transform, it naturally decomposes a graph signal into different frequency components with different variabilities. The Laplacian form of the brain connectivity matrix constructed using Gaussian radial basis function (RBF) kernel plays a major role in graph Fourier transform. Changes in the representation matrix lead to variants of GFT such as alternative graph Fourier transform \cite{sandryhaila2013discrete}, which used adjacency matrix as the shift operator and conducted Jordan decomposition accordingly. Generally speaking, learning graph from data samples is an ill-posed problem. Kernel based methods exhibit well-demonstrated performance in various nonlinear learning tasks in machine learning and signal processing, most of which rely on a preselected kernel presuming task-specific prior information. However, prior information is usually not available due to the complexity of the brain. To overcome this issue, we apply a data-driven graph Laplacian learning framework \cite{dong2016learning} that can effectively construct a brain connectivity network. Afterwards, we use the graph Fourier transform to carry out frequency components partition and further analyze the mechanisms of brain maturation. The whole process is called a graph Laplacian learning Fourier transform (GLFT) to distinguish the graph Laplacian matrix from the one used in graph Fourier transform.

     A similar and well-known method is principal component analysis (PCA), which has been used to extract intrinsic structure in the data by eigen analysis of the covariance matrix \cite{abdi2010principal}. Other matrix decomposition methods such as non-negative matrix factorization \cite{anderson2014non} are also used to extract multiple potentially overlapping brain networks by decomposing data matrices into linear combinations of nonnegative basis functions. A major concern about these methods is the assumption that the data come from a linear space, which is not always true in general cases.  The limitations of the conventional signal processing techniques in irregular graph data such as the lack of a shift-invariance notion have been well discussed in \cite{shuman2013emerging}. Brain networks are invariably complex, which should be characterized using more sophisticated network methods \cite{rubinov2010complex} and therefore graph signal processing is a right tool. \par
As stated previously, it would be interesting to construct the brain network and extract the core regions changing over time to study the mechanism of brain maturation. In this paper, we treat brain imaging data as signals defined on the graph, followed by GLFT process to detect hub regions in a given frequency range. The model is finally applied on data from the Philadelphia Neurodevelopmental Cohort (PNC)\cite{satterthwaite2014neuroimaging}, a large-scale collaborative study from the dbGap database, in which volunteers are healthy adolescents of age 8 to 22 years. The database contains rs-fMRI of nearly 900 subjects and tfMRI of nearly 700 subjects and over 900 subjects for emotion identification and working memory task, respectively.\par
The rest of the paper is organized as follows: in the methodology section, we present in detail the way to construct the brain network and how it is linked to graph signals. We further introduce the GLFT process and explain how it works on the graph signals. The model is then evaluated on both synthetic and real data analysis in the result section, followed by some discussions and concluding remarks.
%Notations: Throughout this paper, uppercase boldface, lowercase boldface, and normal italic letters are used to denote matrices, vectors, and scalars, respectively. The superscript T denotes the transpose of a vector or a matrix. For a matrix A, we denote its i-th row, j-th column, (i, j)-th entry, and trace

\section{Methodology}
\begin{figure*}[htbp]
    \centering
    \includegraphics[width=2\columnwidth]{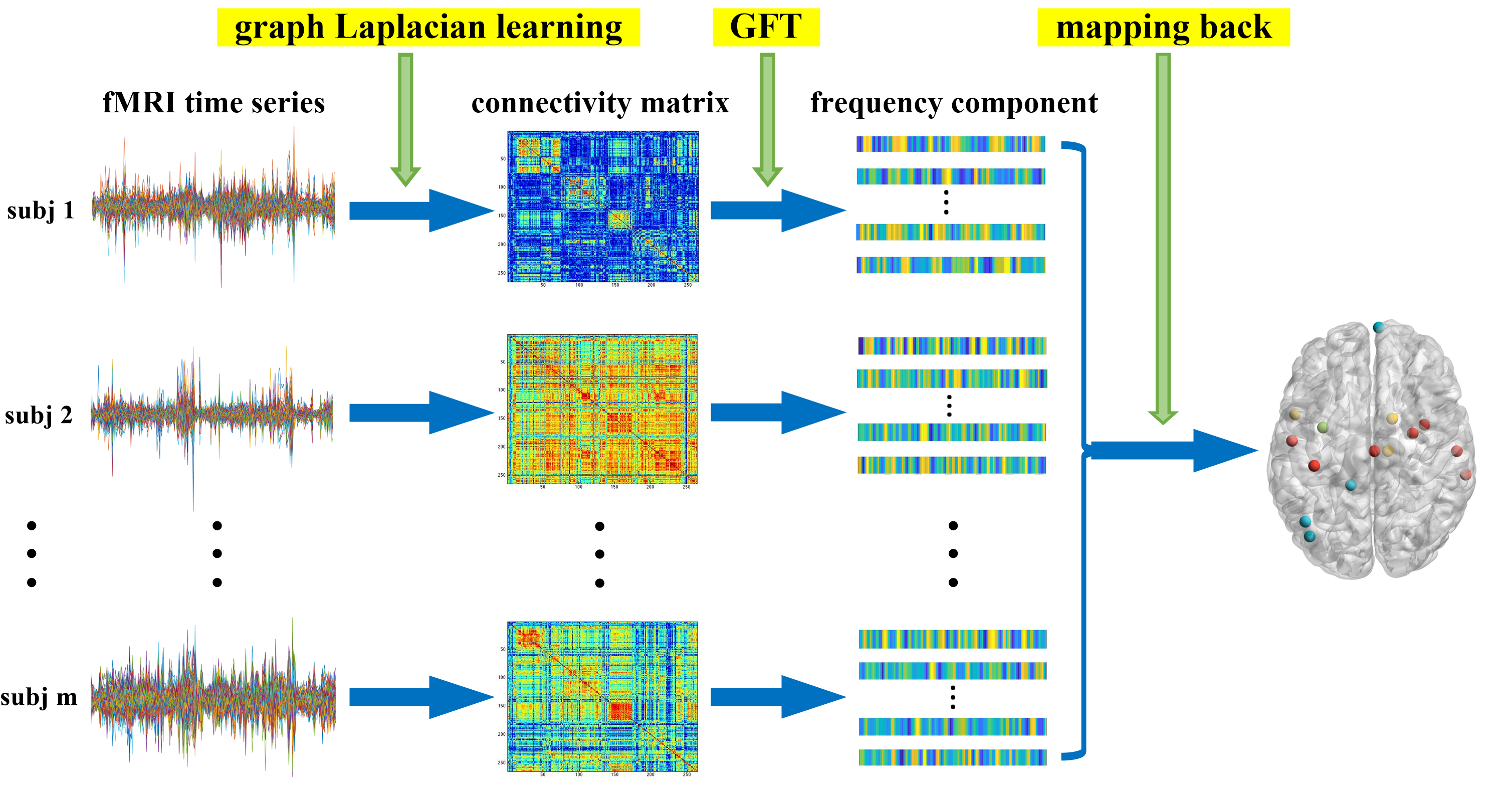}
    \caption{Flow diagram of GLFT process for hubs detection. We treat the fMRI observations as graph signals and construct the connectivity network individually using graph Laplacian learning method. Graph signal was transformed and split into different frequency components using graph Fourier transform and graph filtering. Finally, we report and visualize the selected hub regions from the low frequency components. }
    \label{figure_flow_diagram}
\end{figure*}

In this section, we give a detailed explanation for each step in GLFT process in Fig.\ref{figure_flow_diagram}. First, we introduce the notions of brain network and graph signals to represent fMRI observations. Then, the graph Laplacian matrix is learned as brain connectivity matrix from fMRI observations. Finally, we conduct graph Fourier transform and graph filtering to manipulate the signals into a variety of frequency components.
\subsection{Signal and graph concepts}
We begin by formulating the framework of describing signals defined on a graph. A graph $G(V,E,W)$ with $n$ nodes includes three components: the vertex set $V(|V|=n)$, the edge set $E$ and the weighted adjacency matrix $W$. In brain network analysis, nodes are often defined from a given list of regions of interest (ROI, based on brain atlases \cite{power2011functional}), data-driven clusters (e.g. using independent component analysis (ICA)\cite{allen2011baseline}) or even voxels; edges are often characterized by statistical dependencies between pairs of nodes (or, more generally, groups of nodes) and the weighted adjacency matrix $W$ represents the affinity between the two nodes. Let us assume we have $m\in\mathbb{N}$ subjects. For each subject, we have BOLD time series of $p$ time points and $n$ ROIs available. Let $W=[W_{ij}]\in\mathbb{R}^{n\times{n}}$ denote the association or similarity matrix for a subject. A signal $x$ defined on the vertex set of the graph G is a function from $V \rightarrow \mathbb{R}$, which is often represented by a vector in $\mathbb{R}^n$. The $i$-th element of $x$ represents the real value assigned to the node $i$, which reflects the functional activity of the corresponding brain region.\par
Apart from $W$, we introduce the graph Laplacian $L$ as another important graph associated matrix.
\begin{equation} \label{equ:L definition}
L = D - W,
\end{equation}
where $D$ is referred to degree matrix, which is diagonal with $i$-th element on the diagonal being the sum of $i$-th row in the weight matrix $W$. The graph Laplacian $L$ has following properties: $L$ is a real symmetric matrix and therefore has a complete set of orthonormal eigenvectors $\left\{ f_{i} \right\}_{(i=1,...,n)}$. If $G$ is connected, the associated eigenvalues are real non-negative with the unique smallest eigenvalue being zero \cite{chung1997spectral} $0=\lambda_1<\lambda_2\leq\lambda_3\leq...\leq\lambda_{n}$. By eigendecomposition, the graph Laplacian $L$ can be rewritten as $L=F\Lambda{F^T}$, where $\Lambda$ is diagonal with $\Lambda_{ii}=\lambda_i$ being the $i$-th eigenvalue of $L$ and the $i$-th column of $F$, $f_i$ is the associated eigenvector \cite{hu2013matched}.\par
To investigate the behavior of the graph signal defined on the network, we define the smoothness of $x$ with respect to the Laplacian as
 \begin{equation}
 TV(x):=x^TLx = \frac{1}{2} \Sigma_{i,j=1}^{n}{W_{ij}{(x(i)-x(j))}^2},
 \end{equation}
where $x(i)$ represents the $i$-th element in $x$. It reflects the total variation of the graph signal and has a similar view to optimal embedding \cite{shuman2013emerging,belkin2003laplacian}. According to early work by Chung, a reasonable criterion for a $"good"$ graph map is to minimize the objective function \cite{chung1997spectral} \begin{equation}
\Sigma_{i,j}{{(x(i)-x(j))}^{2}W_{ij}}.
\end{equation}
The definition is reasonable because when $W_{ij}$ is large, the values $x(i)$ and $x(j)$ are expected to be as close as possible, since the weight matrix is intended to measure the function connectional similarity between different brain regions. Otherwise when $W_{ij}$ is small, the contribution of difference between signal values $x(i)$ and $x(j)$ have little effect on the total variation. Therefore, we can treat a signal with small total variation as a smooth signal over the graph while the signals with large total variations being oscillatory.
\subsection{Graph Laplacian learning}
First, we represent the graph signals statistically using factor analysis model, which can explain observations with a potentially smaller number of unobserved latent variables.
Specifically, the graph signal can be written as follows \cite{dong2016learning}:
\begin{equation}
x = Fh + u_x + \epsilon,
\end{equation}
where $x\in\mathbb{R}^{n}$ represents the observed signal, $h\in\mathbb{R}^{n}$ represents the unobserved latent variable, $u_{x}\in\mathbb{R}^{n}$ stands for the mean of $x$ and $\epsilon$ is the noise term. \par
The main consideration is the introduction of eigenvector matrix $F$ as the representation matrix. On one hand, the eigenvector matrix of the graph Laplacian is directly related to the properties of the graph signals by providing a spectral embedding of the graph vertices. One the other hand, it serves as basis to represent graph signals in the frequency domain. Following the similar framework proposed by Dong \cite{dong2016learning}, we assume that $\epsilon$ and $h$ follow a multivariate Gaussian distribution with zero-mean and covariance $\sigma_{\epsilon}I_n$ and $\Lambda^{\dagger}$ ($\Lambda^{\dagger}$ is the Moore-Penrose pseudoinverse of $\Lambda$), respectively, i.e.,
\begin{align}
& \epsilon \thicksim \mathcal{N}(0,{\sigma_{\epsilon}}^{2}I_n) \\
& h\thicksim\ \mathcal{N}(0,\Lambda^{\dagger}).
\end{align}
Given the above prior, the energy of the signal is concentrated mainly on the first few eigen components such that it promotes the smoothness of the graph signal. With the above assumptions, the probability function of the graph signal $x$ is given as:
\begin{equation} \label{equ: X distri}
x\thicksim\ \mathcal{N}(u_x,F\Lambda^{\dagger}F^T+{\sigma_{\epsilon}}^{2}I_n)).
\end{equation}
\par We then introduce a maximum a posteriori(MAP) estimation to estimate $h$ given graph observation $x$ and prior distribution of $h$. Specifically, by applying Bayes' rule and assuming without loss of generality that $u_x = 0$, the MAP estimation of the latent variable $h$ can be written in the following form \cite{gribonval2011should}:
\begin{equation} \label{equ: MAP}
\begin{aligned}
h_{MAP}(x) :&= \arg\underset{h}\max{\ p(h|x)}
\\&=\arg\underset{h}\max{\ p(x|h)p(h)}
\\&=\arg\underset{h}\min{\ (-\ln{p_{E}(x-Fh)}-\ln{p_{H}(h)})}
\\&=\arg\underset{h}\min{\ (-\ln{e^{\frac{-(x-Fh)^T(x-Fh)}{{\sigma_{\epsilon}}^2}}}-\ln{e^{-h^T\Lambda{h}}})}%{-\ln{e^{\frac{{(x-Fh)^{T}(x-Fh)},{\sigma_{\epsilon}}^{2}}}}-\ln{p_{H}(h)}}
\\&=\arg\underset{h}\min{\ h^T\Lambda{h} + k\|x-Fh\|_2^2},
\end{aligned}
\end{equation}
where $p_{E}(x-Fh)$ and $p_{H}(h)$ denote the probability density function of the noise and the latent variable, respectively. $k={\sigma_{\epsilon}^2}$ is the parameter denoting the variance of the noise. Especially, in a noise free scenario when $x=Fh$, the minimization problem is reduced to:
\begin{equation}
h^T\Lambda{h} = (F^{T}x)^T\Lambda{F^{T}x} = x^TLx.
\end{equation}
Thus, it confirms that the proposed model describes the smooth property of the graph signal $x$. Similar conclusion can be made in a noisy setting, where the main component($Fh$) is smooth on the graph.
\par Using a change of variable $y = Fh$, we solve the problem in (\ref{equ: MAP}) with the following objective function \cite{dong2016learning}:
\begin{equation}
\begin{aligned}
\underset{L\in\mathbb{R}^{n\times{n}},Y\in\mathbb{R}^{n\times{p}}}\min &{\alpha\|X-Y\|_F^2+{tr(Y^TLY)}+\beta\|L\|_F^2}
\\ s.t \ &tr(L) = n,
\\    &L_{ij}=L_{ji}\leq0, (i\neq{j}),
\\  & L\cdot\mathbbold{1}=\mathbbold{0},
\end{aligned}
\end{equation}
where $X\in\mathbb{R}^{n\times{p}}$ contains $p$ observations in columns, and $\alpha,\beta$ are two positive parameters regulating the tradeoff of the noise and penalty term. $tr(\cdot)$ and $\|\cdot\|_F$ denote the trace and Frobenius norm of a matrix and $\mathbbold{0}$ and $\mathbbold{1}$ represent the constant zero and one vectors, respectively. The third one(Frobenius norm of $L$) in the objective function is the penalty term to control the off-diagonal elements in $L$. The trace constraint serves as a normalization factor to avoid trivial solutions and the second and third constraints are based on the properties of the graph Laplacian to ensure the learned $L$ is valid.
\par Since the optimization problem is not jointly convex, we adopted an alternating minimization scheme to fix one variable while solving for the other one at each step, which leads to a local minimum of the solution. Specifically, we first initialize $Y$ as the graph signal $X$ and iteratively solve the following optimization problems:
\begin{equation}
\begin{aligned}
\underset{L}\min\ &{{tr(Y^TLY)}+\beta\|L\|_F^2}
\\ s.t \  &tr(L) = n,
\\    &L_{ij}=L_{ji}\leq0,(i\neq{j}),
\\  & L\cdot\mathbbold{1}=\mathbbold{0},
\end{aligned}
\end{equation}
\begin{equation}
\begin{aligned}
\underset{Y}\min\ &{\alpha\|X-Y\|_F^2+{tr(Y^TLY)}}.
\end{aligned}
\end{equation}
The problems can be effectively solved with CVX \cite{grant2015cvx}.
\subsection{Graph Fourier transform}
With the notion of the signal defined on the graph and its graph Laplacian $L$, the graph Fourier transform(GFT) of $x$ with respect to $L$ is defined as
\begin{equation}
\tilde{x}:=F^{T}x.
\end{equation}
The inverse GFT of $\tilde{x}$ with respect to $L$ is defined as \begin{equation}
x=F\tilde{x}.
\end{equation}
Note that $F$ is orthonormal, i.e., $F^{T}F=I$, and the original signal $x$ can be recovered through inverse GFT of the $\tilde{x}$. Thus $x$ and $\tilde{x}$ form a GFT pair.
\par     There are two explanations linking GFT with conventional Fourier Transform. Mathematically, if the eigenvectors in $F$ are of the form of $f_{k}=[1,e^{-i\frac{2\pi{(k-1)}}{n}},...,e^{-i\frac{2\pi{(k-1)(n-1)}}{n}}]^T$, the GFT is reduced to conventional time domain Fourier transform. Specifically, the eigenvectors of a cycle graph are of such form \cite{huang2016graph}. Therefore, conventional Fourier transform can be viewed as a specific case of GFT. On the other hand, the eigenvalues of the graph Laplacian play a similar role in serving as frequency components compared with conventional Fourier transform. In classical Fourier analysis, the eigenvalues $\left\{ 2\pi{(k-1)} \right\}_{(k=1,...,n)}$ carry a specific notion of frequency \cite{shuman2013emerging}: when $k$ approaches one (low frequencies), the associated complex exponential eigenfunctions oscillate slowly, appearing to be smooth, whereas when $k$ is far from one (high frequencies), the associated eigenfunctions oscillate rapidly. In the graph setting, eigenvalues $\lambda_{k}$ have the similar notion of frequency. In fact, we can define the total variation of the eigenvector $f_{k}$ to measure how much the eigenvector over the graph: \begin{equation}TV(f_k):=f_k^TLf_k = f_k^T\lambda_k{f_k} = \lambda_k.
\end{equation}
\par We can conclude confidentially that when $k$ is close to one, the eigenvectors $f_k$ change slowly over the graph, whereas when $k$ is far away from one, the eigenvectors vary rapidly. This is consistent with the conventional Fourier Transform.\par

\subsection{Graph filtering}
Given the above discussion, we are able to manipulate the transformed signal in the frequency domain by designing filters to extract information at a range of frequencies. Following the standard procedure from signal processing, we define a low-pass filter to keep the transformed signal below a given threshold; a high-pass filter to represent the high frequency components; and a band-pass filter to preserve in between. We begin by splitting the diagonal matrix $\Lambda$ with $\lambda_{l}$ and $\lambda_{m}$  being the threshold to separate low and high frequencies.
\begin{align}
\Lambda = &diag(\lambda_1,\lambda_2,...,\lambda_l,0,...,0)+ \notag
\\&diag(0,...,0,\lambda_{l+1},...,\lambda_m,0,...,0) +       \notag
\\&diag(0,...,0,\lambda_{m+1},...,\lambda_n).
\end{align}
The corresponding graph filter for each frequency range can be defined as:
%\begin{equation*}
\begin{align}
&\mathscr{G}_l= diag(\underbrace{1,1,...,1}_{l},0,...,0)\\
&\mathscr{G}_m=diag(\underbrace{0,...,0}_{l},\underbrace{1,...,1}_{m-l},0,...,0)\\
&\mathscr{G}_h=diag(0,...,0,\underbrace{1,...,1}_{n-m}).
\end{align}
%\end{equation*}
\par The effect of the filtering on the graph spectral coefficients is $\mathscr{G}\tilde{x}$. Thus, the graph signal in the frequency domain can be decomposed into low frequency, intermediate frequency and high frequency components $\tilde{x}=\tilde{x}_l+\tilde{x}_m+\tilde{x}_h$. Based on the definition of the inverse Graph Fourier transform, the original graph signal can be reconstructed by $x=x_l+x_m+x_h$ with $x_l,x_m,x_h$ being the inverse signal corresponding to the specific frequencies.
\par It has long been understood that the brain combines certain degree of disorganized behavior with some degree of regularity to form a high complexity system with order and disorder coexisting \cite{sporns2010networks}. In our framework, the ordered part corresponds to low frequency component with small total variation which is believed to help with the organization of the brain functional activities and the disorder portion stands for the high frequency component with high total variation which is more likely to be the noise. The low variability has proven to be essential in the analysis of neurological disease and behavior and strong contribution in learning process \cite{garrett2012modulation,huang2016graph}. As a consequence, we put our emphasis on the learning of the low frequency subspace in the following section.
%\par Interesting findings have been reported by Hu et al \cite{hu2013matched} about Alzheimer's disease, who analyzed the magnitude of the eigenvectors corresponding to different cortical regions. We further improved the model by taking the projection energy of each projected eigenbasis into consideration. Specifically, the inner products on eigenbasis of the signals in the frequency domain were calculated. The spikes of the energy concentration were considered dominant in the frequency domain, which could be further mapped back to detect important cortical regions.
%\centering
\section{Results}
In this section, we first evaluated the performance of the proposed graph learning method on synthetic data. We further applied GLFT for real brain imaging data analysis and reported some interesting findings regarding the brain development stages during adolescence. Finally, we compared the classification results using learned Laplacian matrix as fingerprints to sperate subjects belonging to different age groups with those using the original graph Fourier transform and alternative graph Fourier transform based ones \cite{shuman2013emerging,sandryhaila2013discrete}. Fig.\ref{figure_flow_diagram} gives the summarization and illustration of the main steps involved in GLFT process to identify hub regions in this work.
\subsection{Synthetic data analysis}
\subsubsection{Data preparation}
We generated two types of artificial graphs with 20 vertices that follow the Erdos-Renyi (ER) model \cite{erdos1960evolution} and the Barabasi-Albert (BA) model \cite{barabasi1999emergence}. Specifically, the ER model in this case generates a graph with probability 0.2 for each possible edge independently. The BA model generates a graph by adding new vertices linking with other existing nodes proportional to the degree distribution, namely, the probability of the new vertex to an existing node is the ratio of the degree of that existing node to the sum of degrees of all the existing nodes. We consider unitary weights in this experiment and calculate the ground truth Laplacian based on (\ref{equ:L definition}). For each graph, we generated 100 signals according to (\ref{equ: X distri}) with $u_x = 0$ and $\sigma_{\epsilon} = 0.5$.
\subsubsection{Result and visualization}
Given the graph signal $x$ defined on each graph, we first applied the proposed graph Laplacian learning method to learn the graph Laplacian matrix. For comparison, we calculated the graph Laplacian by two other methods: i) construct the similarity matrix $W$ using RBF kernel and then calculate $L$ in (\ref{equ:L definition}); ii) construct the similarity matrix $W$ using the Pearson correlation at different vertices of the signal and then  calculate $L$ in (\ref{equ:L definition}). To evaluate the performance of these methods quantitatively, we used precision, recall and F-measure as the evaluation metrics statistically. Specifically, the precision measures the percentage of edges in the learned graph that are present in the ground truth graph. On the contrary, the recall measures the percentage of the edges in the ground truth graph that are also in the learned graph. The F-measure is the harmonic average of the precision and recall that evaluates the overall performance of the recovery of the edge set. We repeated the experiment 20 times randomly and the average results are shown in Table.\ref{table_simulation}. The results demonstrated that the graph Laplacian learning method outperformed the RBF-kernel and correlation based ones for all metrics.This implies that the graph Laplacian learning method is more powerful in extracting significant patterns in the network.
\begin{table}
\caption{Simulation result}
\begin{center}
\begin{tabular}{ p{2cm}p{1.5cm}p{1.5cm}p{1.5cm}  }
 \hline
 \multicolumn{4}{c}{B-A model} \\
 \hline
 learning method& precision &recall&F-measure\\
 \hline
 gl-learning   & 0.9413    &0.9474&   0.9132\\
 RBF&   0.8217  & 0.9974   &0.8984\\
 correlation &0.4870 & 0.5211&  0.5031\\

 \hline
\end{tabular}

\begin{tabular}{ p{2cm}p{1.5cm}p{1.5cm}p{1.5cm}  }
 \hline
 \multicolumn{4}{c}{E-R model} \\
 \hline
 learning method& precision &recall&F-measure\\
 \hline
 gl-learning   & 0.6669    &0.8363&   0.7375\\
 RBF&   0.5161  & 0.7755   &0.6761\\
 correlation &0.3336 & 0.3423&  0.3378\\

 \hline
\end{tabular}
\end{center}
\label{table_simulation}
\end{table}

\begin{figure}[htbp]
    \centering
    \includegraphics[width=0.5\textwidth]{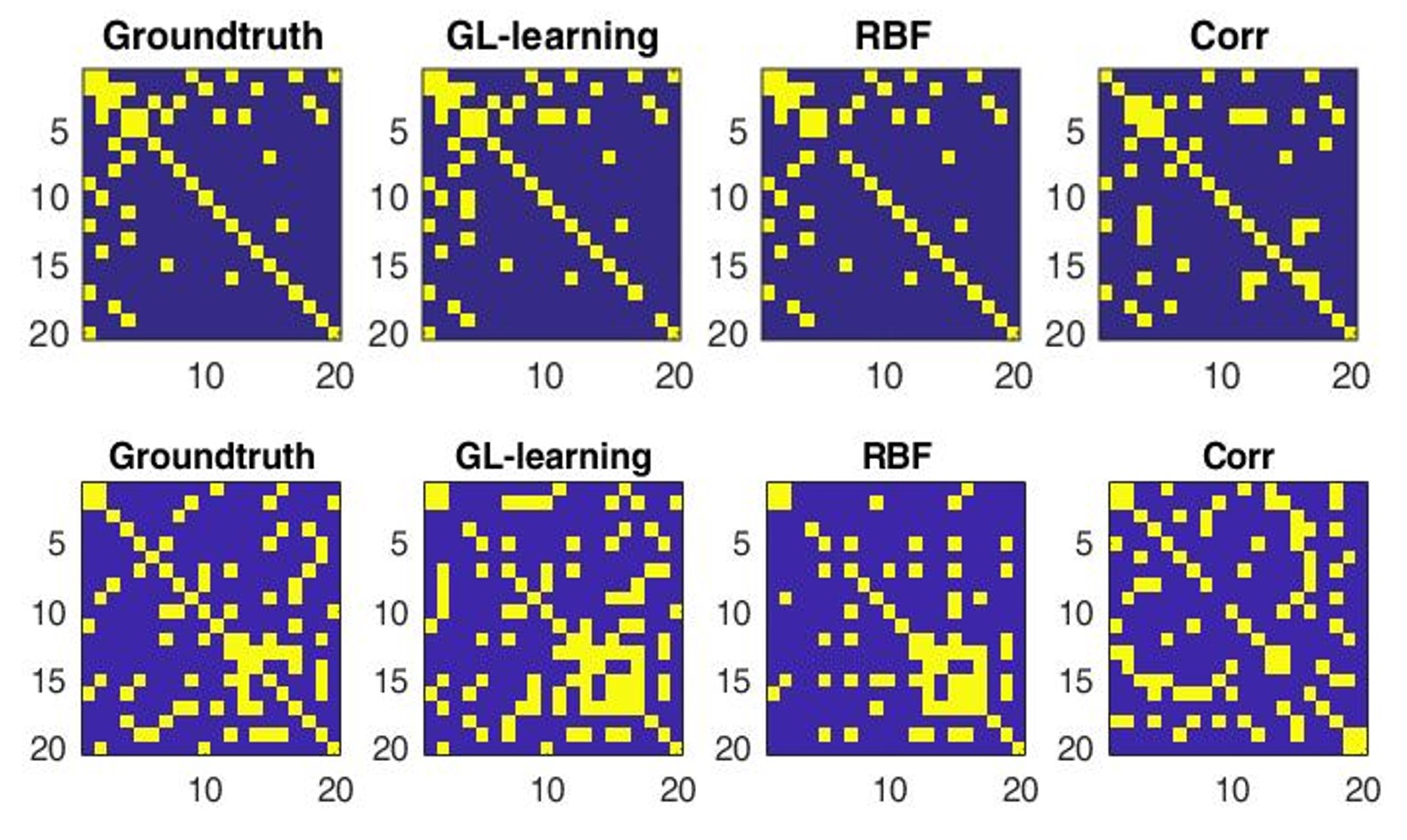}
    \caption{The visualization of the learned graph Laplacian matrix for both E-R and B-A model. The superior part is the simulation from B-A model and the inferior part is the E-R model. From left to right is the groundtruth Laplacian matrix and Laplacian matrix recovered from graph signals learned by GL-learning method, radial basis function kernel and correlation matrix.
}
    \label{figureBA-ER}
\end{figure}

For clarity, we also visualized the learned graph Laplacian matrix on both BA and ER graph in Fig.\ref{figureBA-ER}.
%\vspace{-2em}
\subsection{Real data analysis}
We further applied the GLFT to real neuroimaging data analysis. Specifically, we consider fMRI data collected from Philadelphia Neurodevelopmental Cohort (PNC) \cite{sandryhaila2013discrete} to study the brain maturation process. MRI examinations were conducted on a single 3T Siemens TIM Trio whole-body scanner. Both resting-state and task-based images were collected using a single-shot, interleaved multi-slice, gradient-echo, echo planar imaging sequence.
\subsubsection{Analysis of resting-state MRI data}
\paragraph{Data acquisition and preprocessing}
During the resting-state scan, a fixation cross was displayed as images were acquired. Subjects were instructed to stay awake, keep their eyes open, fixate on the displayed crosshair, and remain still. Total resting state scan duration was 6.2 min \cite{satterthwaite2014neuroimaging}.
\par Standard preprocessing steps were applied using SPM12, including motion correction, spatial normalization to standard MNI space (adult template) and spatial smoothing with a 3mm FWHM Gaussian kernel. The influence of motion (6 parameters) was further addressed using a regression procedure, and the functional time series were band-pass filtered using a 0.01Hz to 0.1Hz frequency range. Finally, we reduced the dimension of the data using the 264 ROI atlas defined by Power et al \cite{power2011functional} with a sphere radius parameter of 5mm. The time sequences of all voxels in the same brain region were averaged so that the data reduced to a $264\times{p}$ matrix for each subject with $p = 124$ denoting the number of time points for each rs-fMRI scan \cite{zille2018fused}. We normalized the regional mean BOLD observation at each time point to maintain consistency of the total energy to avoid extreme spikes due to head motion or drift artifacts in fMRI \cite{huang2016graph}
\par
\begin{figure}[htbp]
    \centering
    \includegraphics[width=0.48\textwidth]{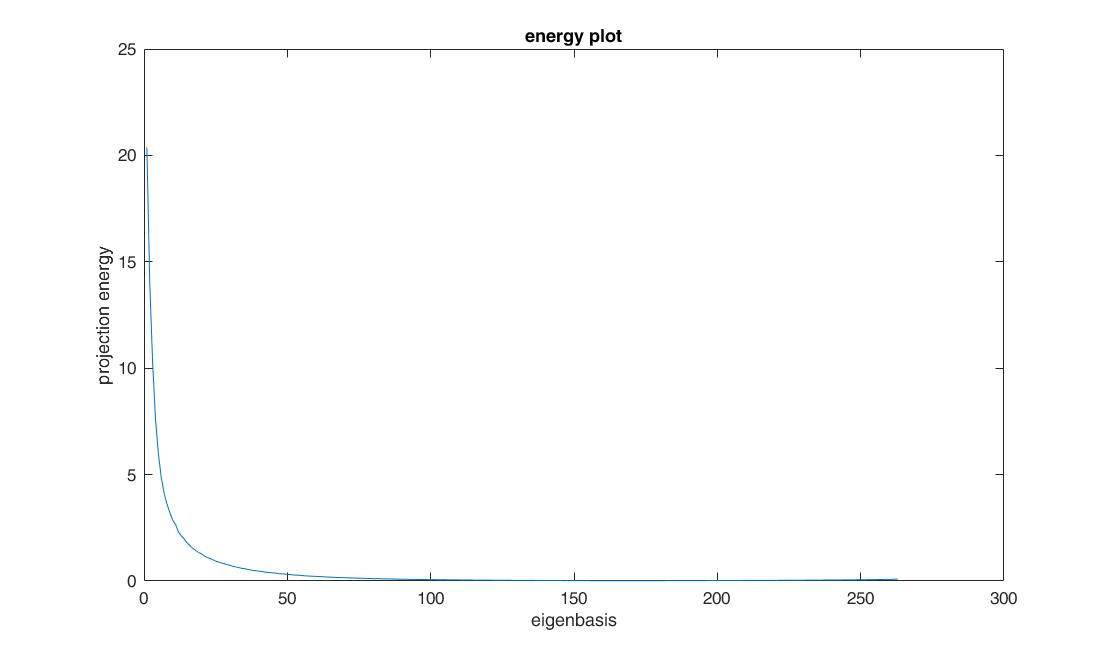}
    \caption{The energy plot of the signal after GLFT process. We calculate the projection energy for all the subjects. We average the projection power for all frequency components except for the first one, which has constant values in its eigenbasis. We plot the energy map toward frequency components in an increasing order. The most concentration of the energy is located in the low frequency components as shown in the figure.}
    \label{energy_plot}
\end{figure}

\begin{figure}[htbp]
    \centering
    \includegraphics[width=0.48\textwidth]{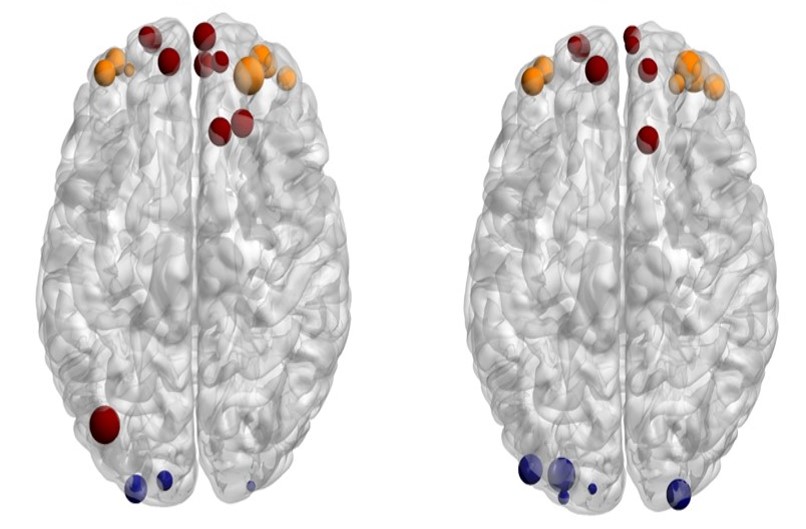}
\caption{The sketch map of the core brain regions coalitions from the rs-fMRI data. The regions in the left part are detected from the children group while the regions in the right part from the young adult group. Most of the regions detected come from Default Mode Network. The size of the nodes represents the detection significance of the region. The visualization of the brain network is by BrainNet Viewer developed by \cite{xia2013brainnet}.
}
    \label{rest_regions}
\end{figure}

 Since we are interested in brain development as a function of age, we select a subset of the full dataset based on age in months. More precisely, each subject whose age is over 18 years is considered to belong to the first class (young adult, age 18-22.58 years, 119 females out of 198 subjects), while subjects whose age is under 12 years belong to the second class (children, age 8.58-12 years, 108 females out of 191 subjects).
\paragraph{Result and visualization}
We followed the procedures illustrated in Fig.\ref{figure_flow_diagram} to identify hub regions for each age group. Specifically, we treated fMRI observations of each time point on different brain regions as graph signals defined on the brain network and applied the graph Laplacian learning algorithm to estimate the graph Laplacian for each subject. Then we conducted graph Fourier transform and graph filtering to decompose the graph signal into low, intermediate and high frequency components. Hu et al \cite{hu2013matched} found
interesting results by analyzing the magnitude of the eigenvectors corresponding to different cortical regions and showed differences in Alzheimer's disease. We further improved the model by taking the projection energy of each projected eigenbasis into consideration. The spikes of the energy concentration were considered dominant in the frequency domain, which could be further mapped back to detect important cortical regions. Specifically, the inner products on eigenbasis of the signals in the low frequency domain were calculated. The core regions were detected whose averaged magnitude of eigenvectors were two standard deviations higher than mean value \cite{bassett2008hierarchical} with cutoffs splitting the frequency ranges selected using cross-validation, i.e, $l = 19$. We plotted the energy concentration of the transformed signals as a reason why we put our emphasis on low frequency part in Fig.\ref{energy_plot}. The plot suggested that the majority of the information contained in the low frequency part. We explored group differences in the organization of the networks at a regional level by statistical testing for a significant between-group difference in the detected hubs and reported the result in Table.\ref{table:rest_you} and Table.\ref{table:rest_old} for children and young adult group, respectively. Regions with * in the table have been reported previously and the reference behind the regions point to the corresponding literature with 10 out of 13 regions in both age stages. We also visualized the locations of the detected brain regions in Fig.\ref{rest_regions}. When we compare the difference between age groups, a significant number of common regions were observed, most of which are listed in the default mode network (DMN) \cite{bucknerbrain}. This finding verifies our previous hypothesis that the low frequency segment represents the ordered regularity part and is consistent with the fact that DMN is dominant at resting state.
\begin{table}
\begin{center}
\caption{Cortical hubs estimated from children stage rs-fMRI}
  \begin{tabulary}{0.48\textwidth}{LCL}
         Location & Statistic test(P-value)                                                  \\
    \hline
Left Cerebrum, Parietal Lobe, Precuneus*\cite{grady2006age} &  $2.48e^{-9}$  \\
Left Cerebrum, Limbic Lobe, Cingulate Gyrus &  $2.10e^{-13}$ \\
Left Cerebrum, Limbic Lobe, Posterior Cingulate*\cite{reyngoudt2012age} & $8.28e^{-5}$\\
Right Cerebrum, Frontal Lobe, Superior Frontal Gyrus*\cite{gogtay2004dynamic} & $8.84e^{-7}$ \\
Right Cerebrum, Temporal Lobe, Inferior Temporal Gyrus*\cite{gogtay2004dynamic} & $4.73e^{-10}$ \\
Right Cerebrum, Temporal Lobe, Middle Temporal Gyrus*\cite{iidaka2002age} & $4.07e^{-4}$\\
Left Cerebrum, Temporal Lobe, Middle Temporal Gyrus*\cite{gogtay2004dynamic} & $1.62e^{-5}$ \\
Right Cerebrum, Occipital Lobe, Cuneus*\cite{garrett2012modulation} & 0.53 \\
Right Cerebrum, Frontal Lobe, Middle Frontal Gyrus*\cite{gogtay2004dynamic} & $8.70e^{-11}$ \\
Right Cerebrum, Parietal Lobe, Inferior Parietal Lobule*\cite{chi1977gyral} & 0.1398 \\
Left Cerebrum, Frontal Lobe, Medial Frontal Gyrus*\cite{rubia2000functional} & 0.0342 \\
Right Cerebrum, Sub-lobar, Extra-Nuclear & $1.67e^{-14}$ \\
Right Cerebrum, Temporal Lobe, Superior Temporal Gyrus*\cite{gogtay2004dynamic} & $1.25e^{-4}$ \\

\end{tabulary}
\label{table:rest_you}
\end{center}

\end{table}
\begin{table}
\caption{Cortical hubs estimated from young adult stage rs-fMRI}
\begin{center}
  \begin{tabulary}{0.48\textwidth}{LCL}
         Location & Statistic test(P-value)                                                  \\
    \hline
Left Cerebrum, Parietal Lobe, Precuneus*\cite{grady2006age} &  $2.48e^{-9}$  \\
Left Cerebrum, Limbic Lobe, Cingulate Gyrus &  $2.10e^{-13}$ \\
Left Cerebrum, Limbic Lobe, Posterior Cingulate*\cite{reyngoudt2012age} & $8.28e^{-5}$\\
Left Cerebrum, Frontal Lobe, Medial Frontal Gyrus*\cite{rubia2000functional} & 0.0342 \\
Right Cerebrum, Frontal Lobe, Middle Frontal Gyrus*\cite{gogtay2004dynamic} & $8.70e^{-11}$ \\
Right Cerebrum, Frontal Lobe, Superior Frontal Gyrus*\cite{gogtay2004dynamic} & $8.84e^{-7}$ \\
Right Cerebrum, Temporal Lobe, Inferior Temporal Gyrus*\cite{gogtay2004dynamic} & $4.73e^{-10}$ \\
Right Cerebrum, Temporal Lobe, Middle Temporal Gyrus*\cite{iidaka2002age} & $4.07e^{-4}$\\
Left Cerebrum, Temporal Lobe, Middle Temporal Gyrus*\cite{gogtay2004dynamic} & $1.62e^{-5}$ \\
Right Cerebrum, Occipital Lobe, Cuneus*\cite{garrett2012modulation} & 0.53 \\
Right Cerebrum, Parietal Lobe, Inferior Parietal Lobule*\cite{chi1977gyral} & 0.1398 \\
Right Cerebrum, Sub-lobar, Extra-Nuclear & $1.67e^{-14}$ \\
Right Cerebrum, Temporal Lobe, Superior Temporal Gyrus*\cite{gogtay2004dynamic} & $1.25e^{-4}$ \\

\end{tabulary}
\end{center}
\label{table:rest_old}
\end{table}

\subsubsection{Analysis of task-state MRI data}
\paragraph{Data acquisition and preprocessing}
In our work, we applied our method on two types of task-state MRI data from PNC \cite{satterthwaite2016philadelphia} to understand how brain maturation mediates cognitive development. The task involves emotion identification and working memory examination and detailed conduction of the experiments can be found in the paper by Satterthwaite et al \cite{satterthwaite2014neuroimaging}.
\par The emotion identification employs a fast event-related design with a jittered inter-stimulus interval(ISI). 60 faces in total were displayed with neutral, happy, sad, angry or fearful expressions performed by volunteered actors(50\% female) in color photographs. The actors were trained by professional directors to express a range of facial expressions and a subset of expressions was kept based on high degree of accurate identification(80\%) by raters. Participants were required to label the emotion corresponding to the expressions displayed. Each face was displayed for 5.5 s followed by a variable ISI from 0.5s to 18.5 s, during which a complex crosshair (that matched the faces' perceptual qualities) was displayed. Total emotion identification task duration was 10.5 min \cite{satterthwaite2014neuroimaging}.
\par The working memory studies were performed via a fractal version of the standard n-back task, which was proved to be a reliable indicator of the executive system and was able to avoid lexical processing abilities that evolve during development \cite{brown2004developmental,schlaggar2002functional}. The task was conducted by presenting complex geometric figures (fractals) for 500 ms, followed by a fixed ISI of 2500 ms. There are three levels of test in total: 0-back, 1-back and 2-back. In 0-back setting, participants responded to a specified target fractal; in 1-back setting, participants responded if the current fractal was identical to the previous one; in 2-back setting, participants responded if the current fractal was identical to the previous two trails. Total working memory task duration was 11.6 min \cite{satterthwaite2014neuroimaging}.
\par The preprocessing steps for task state fMRI data were similar to that adopted for resting state fMRI data, detailed in Result section B.1.a. The only differences were at the final step that the time sequences reduced to a $264\times p$ matrix for every subject with $p$ = 210 and $p$ = 231 denoting the number of time points for emotion identification and working memory tasks, respectively, due to the total duration of the MRI scan. The same manner was adopted to partition the age groups, with 154 and 204 subjects belonging to the younger group and 148 and 207 subjects belonging to the older group for emotion and working memory task, respectively.
\paragraph{Result and visualization}
\begin{figure}
    \centering
    \includegraphics[width=0.5\textwidth]{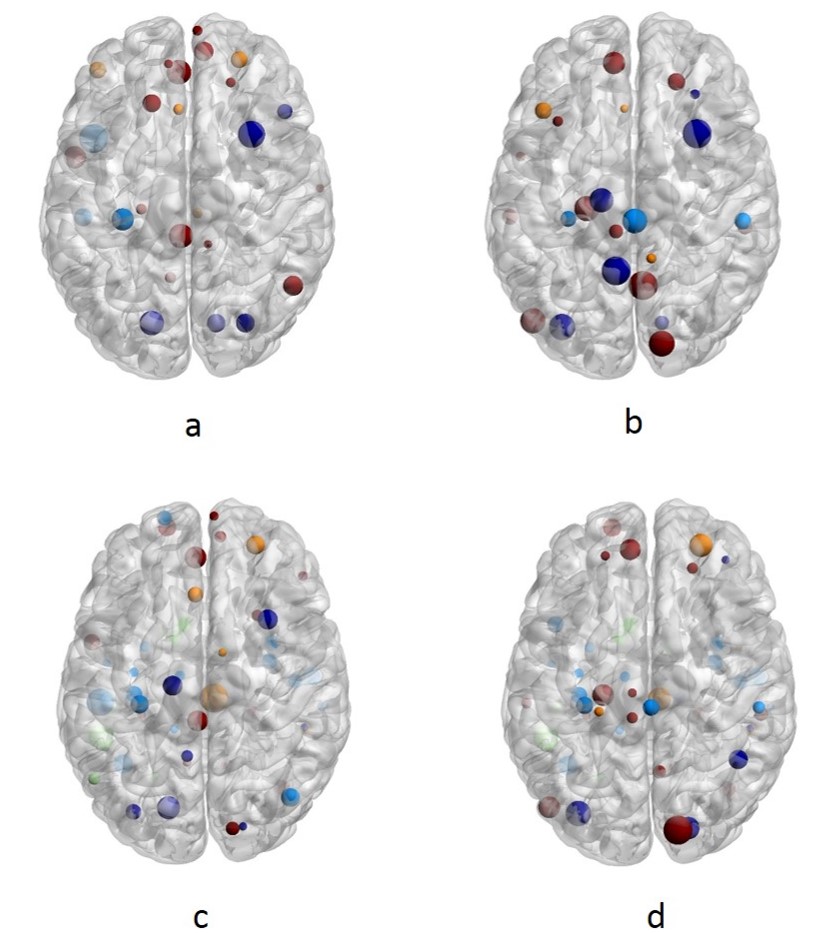}
    \caption{The sketch map of the core brain regions coalitions identified from the tfMRI data. The superior part (a,b) indicates the network constructed from emotion identification task while the inferior part (c,d) indicates the working memory task based network. The regions in left part are identified from children group and the regions in right part from the young adult group. The size of the nodes represents the detection significance of the region.
}
    \label{figure2}
\end{figure}
Following the same steps in rs-fMRI study, we display the estimated task related core regions along brain maturation process in Fig.\ref{figure2} using the BrainNet Viewer \cite{xia2013brainnet}. with thresholds being $l=45$ and $l=40$ for working memory task and emotion identification task, respectively. Different from rs-fMRI study, task-based age related regions are observed for both emotion and working memory task such as postcentral parietal gyrus and parahippocampal gyrus (part of cerebral cortex, closely related to hippocampus). The result is within our expectation since it has been reported that the different activation patterns occur in different adolescence stages facing the same task \cite{iidaka2002age}. To exclude unexpected interference, the joint core regions which play a key role in both cognitive tasks were identified as hubs. We listed these hubs in Table.\ref{table:young} and Table.\ref{table:old} for children and young adult group, respectively. 8 of the 11 hubs in children stage and 6 of 9 hubs in young adult stage have been reported in the literatures \cite{grady2006age,gogtay2004dynamic,iidaka2002age,chi1977gyral,rubia2000functional,grady1995age,shaw2006intellectual} to be closely related to the brain maturation.
\begin{table}
\caption{Cortical hubs estimated from children stage tfMRI}
\begin{center}
  \begin{tabulary}{0.48\textwidth}{LCL}
         Location & Statistic test(P-value)                                         \\
    \hline
    Left Cerebrum, Parietal Lobe, Postcentral Gyrus*\cite{gogtay2004dynamic}
       &  $6.14e^{-5}$  \\
    Right Cerebrum, Frontal Lobe, Medial Frontal Gyrus*\cite{iidaka2002age}
. &  $2.87e^{-6}$ \\
Right Cerebrum, Parietal Lobe, Inferior Parietal Lobule*\cite{chi1977gyral} & $6.52e^{-5}$\\
Left Cerebrum, Sub-lobar, Insula & 0.0078 \\
Right Cerebrum, Parietal Lobe, Precuneus & $1.55e^{-5}$ \\
Right Cerebrum, Frontal Lobe, Superior Frontal Gyrus*\cite{gogtay2004dynamic} & 0.1492 \\
Left Cerebrum, Frontal Lobe, Medial Frontal Gyrus*\cite{rubia2000functional} & 0.0438 \\
Left Cerebrum, Occipital Lobe, Middle Occipital Gyrus & 0.0826 \\
Left Cerebrum, Limbic Lobe, Parahippocampal Gyrus*\cite{grady1995age} & 0.9779 \\
Right Cerebrum, Temporal Lobe, Middle Temporal Gyrus*\cite{iidaka2002age} & 0.0654\\
Left Cerebrum, Temporal Lobe, Middle Temporal Gyrus*\cite{gogtay2004dynamic} & $1.87e^{-5}$\\

  \end{tabulary}

\end{center}
\label{table:young}
\end{table}
%\boxtimes

\begin{table}
\caption{Cortical hubs estimated from young adult stage tfMRI}
\begin{center}
  \begin{tabulary}{0.48\textwidth}{LCL}
         Location & Statistic test(P-value)                                                  \\
    \hline
  Left Cerebrum, Temporal Lobe, Superior Temporal Gyrus*\cite{gogtay2004dynamic} &  0.0015  \\
  Left Cerebrum, Sub-lobar, Insula & 0.0078 \\
Left Cerebrum, Limbic Lobe, Cingulate Gyrus & $2.65e^{-4}$ \\
Right Cerebrum, Frontal Lobe, Medial Frontal Gyrus*\cite{shaw2006intellectual} & $1.25e^{-4}$\\
Left Cerebrum, Frontal Lobe, Medial Frontal Gyrus*\cite{rubia2000functional} & 0.0272 \\
Left Cerebrum, Temporal Lobe, Middle Temporal Gyrus*\cite{gogtay2004dynamic} & $5.57e^{-4}$ \\
Left Cerebrum, Parietal Lobe, Precuneus*\cite{grady2006age} & 0.0044 \\
Right Cerebrum, Frontal Lobe, Superior Frontal Gyrus*\cite{gogtay2004dynamic} & 0.3235\\
Right Cerebrum, Sub-lobar, Extra-Nuclear & 0.5862\\
  \end{tabulary}
\end{center}
\label{table:old}
\end{table}
\subsubsection{Brain maturation analysis}
In this section, we summarize the findings and interpret the results with brain maturation. We notice that for resting state fMRI data, a series of common regions are detected for both age groups. These regions have been identified within the default mode network. Recent studies report that the same general pattern for the DMN in adults is also found in children age from 7 to 9 years and 7-12 years \cite{thomason2008default,supekar2010development}. It indicates that the DMN is relatively stable at stages from childhood to puberty to adulthood. Since DMN is dominantly active during the rs-fMRI scan periods, it is reasonable that the detected regions lie mainly in DMN and similar patterns are observed across age groups. Especially we find that superior temporal gyrus (STG) is reported in both age groups, which is commonly believed to be essential during brain maturation process \cite{raichle2015brain,raichle2001default}. In a longitudinal study conducted by Nitin Gogtay and his colleagues \cite{gogtay2004dynamic}, STG is identified with the most variation and speed of gray matter loss(brain maturation) during adolescence stage. \par
In tfMRI analysis, the middle temporal gyrus is observed in the children stage while the superior temporal gyrus is observed in the order one. Both middle and superior temporal gyrus are thought to be heteromodal association sites (along with prefrontal and inferior parietal cortices) and are involved with integration of memory, audiovisual association and object-recognition functions \cite{gogtay2004dynamic}. The findings on the temporal cortex is consistent with the report that the superior temporal cortex, which contains association areas that integrate information from several sensory modalities, matures last \cite{gogtay2004dynamic}. In parietal lobe, the postcentral gyrus is detected in the children stage and has been reported to be the beginning part of the gray matter loss. The inferior parietal cortex has also been reported to mature during puberty. Neuroimaging studies demonstrated that the hippocampus has significant activation during the encoding of faces or pictures in young adults, but the activation is reduced in older adults \cite{grady1995age,grady1999effects}. We also draw the similar conclusion because the hippocampus is detected only in the children group. The frontal lobes that are involved in executive function, attention and motor coordination, are observed for both age groups.
\begin{table}
\caption{Result for fMRI data analysis}
\begin{center}
\begin{tabular}{ p{2cm}p{1.5cm}p{1.5cm}p{1.5cm}  }
%\begin{center}
 \hline
 \multicolumn{4}{c}{rs-fMRI} \\
 \hline
 & GLFT &RBF&Corr\\
 \hline
 ACC   & 0.9306    &0.8870&   0.9279\\
 SEN&   0.9295  & 0.8945   &0.9242\\
 SPE &0.9321 & 0.8803&  0.9318\\
 \hline
 %\end{center}
\end{tabular}
\begin{tabular}{ p{2cm}p{1.5cm}p{1.5cm}p{1.5cm}  }
 \hline
 \multicolumn{4}{c}{emoid-fMRI} \\
 \hline
 & GLFT &RBF&Corr\\
 \hline
 ACC   & 0.9569    &0.9110&   0.9469\\
 SEN&   0.9595  & 0.9262   &0.9462\\
 SPE &0.9542 & 0.8963&  0.9475\\
 \hline
\end{tabular}
\begin{tabular}{ p{2cm}p{1.5cm}p{1.5cm}p{1.5cm}  }
 \hline
 \multicolumn{4}{c}{nback-fMRI} \\
 \hline
 & GLFT &RBF&Corr\\
 \hline
 ACC   & 0.9392    &0.8856&   0.9417\\
 SEN&   0.9271  & 0.9129   &0.9271\\
 SPE &0.9514 & 0.8586&  0.9564\\
 \hline
\end{tabular}
\end{center}

Note.In this table, ACC, SEN and SPE are the abbreviations of accuracy, sensitivity and specificity, respectively.
\label{wenxing}
\end{table}
For further validation of the result, we separated the age groups using the learned brain network via linear SVM (support vector machine) classifier \cite{scholkopf2002learning}. We also constructed the brain network using RBF kernel and correlation matrix to compare the performance. Sensitivity, specificity and accuracy are used as metrics to evaluate the performance. Specifically, sensitivity measures the percentage of children that are correctly identified in the younger group and the specificity measures the proportion of actual young adults that are correctly identified in the older group. Accuracy evaluates the overall performance of the classifier by indicating the percentage of the correctly identified subjects. 10-fold cross-validation was carried out to avoid overfitting and we list the result in the Table.\ref{wenxing}.
\par In general, the performance of GLFT method is better than RBF and correlation based one(except for the slightly worse result compared with correlation in working memory task). It demonstrates that the proposed method is comparably powerful in extracting the age-related features. The high accuracies in distinguishing age groups also suggest the disparity of brain maturation stages, which implies the changes and transformations of the brain network during puberty.
\section{Conclusion}
We analyzed the brain functional connectivity differences between children and young adults using GLFT to explore the mechanisms of brain maturation during adolescence. Specifically, neuroimaging data were parcellated into regions of interest according to the pre-defined functional atlas \cite{power2011functional}. The fMRI observations were viewed in a graph setting to become graph signals defined on the vertex set or namely, different functional brain regions. Given these graph signals, the connectivity matrix of the pairwise regions were estimated via a graph Laplacian learning algorithm. These signals were further transformed into the frequency domain, manipulated into several frequency ranges, analyzed in the filtered frequency components, and recovered from each frequency portion using graph signal processing tool. Compared with the graph Fourier transform and its variants, the GLFT can facilitate the extraction of the brain connectivity network directly from the BOLD observations based on the smoothness conditions on the graph. In both real and synthetic data analysis, we show that the proposed learning method outperforms the GFT and Alternative GFT based one. Based on PNC dataset containing both task and task-free fMRI imaging, the core regions corresponding to the brain development at normally developing adolescent stage were identified through spectral analysis of the graph Laplacian eigenfunctions. The findings in this experiment are consistent with the original hypothesis from brain network analysis \cite{sporns2010networks}. The hub regions reported are also consistent with the longitudinal biological experiment on the dynamic map of gray matter loss(brain maturation) during puberty \cite{gogtay2004dynamic}. \par
In conclusion, we present a GLFT approach to study the brain functional difference among adolescent age groups and demonstrate it on a real large dataset. The core regions reported in this paper can serve as biomarkers for brain maturation at youth stage. We hope this model can find other applications in neuroimaging.
%\appendices
%% you can choose not to have a title for an appendix
%% if you want by leaving the argument blank
%\section{}

\section*{Acknowledgment}

This work has been supported by NIH (R01 GM109068, R01 MH104680, R01 MH107354, P20 GM103472, R01 REB020407, 1R01 EB006841) and NSF (\#1539067).

\ifCLASSOPTIONcaptionsoff
  \newpage
\fi


\bibliographystyle{IEEEtran}
\bibliography{Wang}
%\vfill
\end{document}